# Automotive HSMs: Architectural Challenges and Security Implications

**Krishna Teja Medam and Austin Bruce**
Vehicle Cyber Engineering, University of Detroit Mercy, Detroit, United States of America

Corresponding authors: Krishna Teja Medam (e-mail: kmedam@udmercy.edu) and Austin Bruce (e-mail: abruce@udmercy.edu).

**ABSTRACT** Automotive electronic control units (ECUs) increasingly depend on hardware-rooted security to protect software integrity, authenticity, and lifecycle management in the presence of remote and physical threats. Hardware Security Modules (HSMs) have become a key building block in automotive system-on-chips (SoCs), providing isolated cryptographic services, secure key storage, and controlled execution under stringent real-time and cost constraints. This paper presents an architectural analysis of automotive HSMs and examines their role in establishing secure boot and hardware roots of trust. We first survey common HSM integration models used in production ECUs and discuss their flexibility and current automotive use cases. We then introduce realistic threat models to motivate hardware-backed security controls and analyze how HSM design choices influence secure boot chains of trust, secure storage, secure execution, and software signing mechanisms. Key tradeoffs between isolation, performance, updateability, and attack surface are discussed, with optional consideration of side-channel implications. The paper concludes by highlighting open challenges and future directions for scalable and resilient automotive hardware security. Finally, we discuss emerging challenges such as cryptographic agility and post-quantum readiness that are likely to shape the next generation of automotive HSM architectures.

**INDEX TERMS** Automotive cybersecurity, hardware security module (HSM), secure boot, hardware root of trust, secure key management, embedded system security, automotive system-on-chip (SoC).

## I. INTRODUCTION

### A. Motivation: Software-Defined Vehicles and Expanding ECU Attack Surface

The automotive industry is undergoing a fundamental transformation toward software-defined vehicles, where functionality, performance, and compliance are increasingly realized through software rather than fixed hardware logic. Modern vehicles integrate a large number of electronic control units (ECUs) connected via heterogeneous in-vehicle networks and exposed through external interfaces such as diagnostics, telematics, and over-the-air (OTA) updates. While this evolution enables faster feature deployment and lifecycle maintenance, it also introduces new attack vectors that significantly expand the vehicle's cyber-physical attack surface [1], [2].

Recent studies have demonstrated that vulnerabilities in firmware, communication stacks, and update mechanisms can be exploited to gain persistent control over ECUs, including those responsible for safety-critical functions. As vehicles become more connected and software-centric, ensuring the authenticity and integrity of software throughout the vehicle lifecycle has become a foundational security requirement rather than an optional enhancement [3].

### B. Limitations of Software-Only Security Mechanisms

Software-only security approaches, including application-layer cryptography and operating system isolation, provide limited protection against realistic automotive adversaries. Attackers with access to diagnostic interfaces, firmware update channels, or physical ECUs can bypass software controls through memory extraction, firmware modification, or manipulation of early boot stages. Prior work has shown that without hardware-enforced trust anchors, secure boot mechanisms can be subverted by replacing bootloaders or extracting cryptographic keys from non-volatile memory [4].

Moreover, automotive ECUs operate under stringent real-time constraints and cost limitations, which restrict the complexity and runtime overhead of software-based security solutions. These constraints necessitate security mechanisms that provide strong isolation and tamper resistance while maintaining deterministic behavior and low latency.

### C. Role of Hardware-Rooted Trust in Automotive Systems

Hardware rooted trust establishes a reliable foundation for system security by anchoring critical operations such as boot verification, key protection and security critical execution in immutable or isolated hardware components. In automotive system-on-chips (SoCs), this role is commonly fulfilled by Hardware Security Modules (HSMs), which provide dedicated cryptographic services, secure key storage, and controlled execution environments isolated from application software.

Automotive HSMs are increasingly adopted to support secure boot, firmware authentication, secure diagnostics, and cryptographic services for in-vehicle communication. Unlike general-purpose trusted execution environments or external security modules, automotive HSMs are specifically designed to balance isolation strength, flexibility, and cost within long-lived embedded platforms. Consequently, HSM architecture and integration choices have a direct impact on the robustness and scalability of the vehicle's overall security posture.

### D. Contributions of This Paper

This paper presents an architectural analysis of automotive Hardware Security Modules and examines their role in establishing secure boot and hardware roots of trust in modern automotive SoCs. The key contributions are as follows:

A. A systematic overview of automotive HSM architectures and integration models, highlighting common design patterns and trust boundaries.

B. An analysis of how HSM design choices influence secure boot chains, secure storage, secure execution, and software signing mechanisms, grounded in realistic attacker models.

C. A discussion of architectural tradeoffs involving isolation, performance, updateability, and attack surface, with optional consideration of side-channel implications.

D. Identification of open challenges and future directions, including cryptographic agility and evolving requirements that may shape next-generation automotive HSM designs.

By focusing on architectural considerations rather than cryptographic algorithms, this paper aims to provide practical insights for system architects and security engineers designing scalable and resilient automotive platforms.

## II. BACKGROUND: AUTOMOTIVE HARDWARE SECURITY MODULES

### A. Definition and Objectives of Automotive HSMs

An automotive Hardware Security Module (HSM) is a dedicated hardware component integrated into an automotive system-on-chip (SoC) that provides hardware-enforced security services to the rest of the electronic control unit (ECU). Its primary purpose is to establish a hardware root of trust by securely storing cryptographic secrets, executing security-critical operations in isolation, and enforcing security policies independently of application software [5].

Unlike purely software-based security mechanisms, automotive HSMs rely on physical and logical isolation to protect assets such as cryptographic keys, certificates, and configuration data from compromise, even in the presence of a fully compromised main processor. Typical services provided by an HSM include cryptographic primitives (e.g., symmetric and asymmetric operations), secure key management, random number generation, secure boot support, and authentication services for diagnostics and in-vehicle communication.

The objectives of automotive HSMs extend beyond confidentiality of cryptographic material. They are designed to ensure integrity and authenticity of software during boot and update processes, to support secure lifecycle management across production, deployment, servicing, and decommissioning phases, and to provide a trusted execution environment for security-critical control flows. Importantly, these objectives must be achieved under the strict real-time, safety, and cost constraints characteristic of automotive embedded systems, distinguishing automotive HSMs from general-purpose security modules.

### B. HSMs vs. Alternative Security Primitives (Software Crypto, TPMs, TEEs)

Automotive HSMs coexist with, and are sometimes compared to, several alternative security primitives, each with different trust assumptions and architectural implications.

Software-based cryptography relies on cryptographic libraries executed on the main CPU, often supported by memory protection mechanisms provided by the operating system. While flexible and cost-efficient, software-only approaches offer limited resistance against adversaries with physical access or low-level system privileges. Keys stored in system memory or flash can be extracted, and early boot stages remain vulnerable without a hardware-enforced trust anchor. As a result, software cryptography alone is insufficient to protect against realistic automotive threat models.

Trusted Platform Modules (TPMs) provide a standardized discrete or integrated hardware root of trust widely used in enterprise and PC environments. TPMs offer strong security guarantees for key storage, attestation, and measured boot. However, their command-based interface, standardized firmware stack, and relatively high latency make them less suitable for hard real-time automotive workloads. Additionally, TPM cost, integration complexity, and mismatch with automotive safety and lifecycle requirements have limited their adoption in production ECUs.

Trusted Execution Environments (TEEs), such as ARM TrustZone-based designs, provide logical isolation by partitioning the main processor into secure and non-secure worlds. TEEs enable flexible secure applications and can host complex security services, but they share hardware resources with non-secure software and typically depend on a secure boot chain rooted elsewhere. As a result, TEEs alone do not constitute a hardware root of trust and are often layered on top of an HSM rather than replacing it.

In contrast, automotive HSMs are purpose-built to act as a minimal, highly isolated security anchor tightly integrated into the SoC. They trade general-purpose programmability for reduced attack surface, deterministic behavior, and tighter coupling with boot and key management flows, making them well suited to automotive use cases.

### C. Automotive Constraints Influencing HSM Design

The design of automotive HSMs is shaped by a unique combination of technical, economic, and lifecycle constraints that differ significantly from those of consumer or enterprise security platforms.

Real-time requirements; Automotive ECUs often perform safety-critical control functions with strict timing guarantees. Security services provided by the HSM, such as message authentication, key derivation, or secure boot verification, must therefore exhibit predictable and bounded latency. This requirement influences architectural choices such as limited instruction sets, hardware-accelerated cryptographic primitives, and simple communication interfaces between the main CPU and the HSM. Excessive context switching, complex firmware stacks, or non-deterministic execution models are generally avoided to preserve real-time behavior.

Cost and silicon area; Automotive systems are produced at high volumes and are subject to tight cost targets. As a result, HSMs must provide sufficient security guarantees while minimizing silicon area, power consumption, and integration complexity. This constraint often leads to design tradeoffs, such as limiting memory size, restricting programmability, or sharing certain peripherals with the main SoC. Unlike discrete security modules, automotive HSMs are typically integrated on-die, requiring careful balancing of security isolation against overall SoC cost.

Long vehicle lifecycles; Vehicles are expected to remain operational for 10 to 20 years, during which cryptographic algorithms, threat models, and regulatory requirements may evolve significantly. Automotive HSMs must therefore support long-term maintainability, including secure firmware updates, key revocation, and algorithm agility. At the same time, parts of the HSM (such as boot ROM or root keys) are often immutable after manufacturing, creating tension between strong hardware anchoring and future adaptability. These lifecycle considerations play a critical role in architectural decisions related to updatability, versioning, and trust anchor management.

## III. AUTOMOTIVE HSM ARCHITECTURES AND DESIGN TRADEOFFS

Automotive Hardware Security Modules are realized through a range of architectural integration models that reflect different assumptions about isolation strength, performance requirements, and system cost. While all automotive HSMs aim to provide a hardware root of trust, their internal structure, execution model, and interaction with the host system vary significantly. This section surveys common HSM architectures deployed in automotive SoCs and analyzes the key tradeoffs that influence their security properties and practical deployment.

### *A. Common HSM Integration Models*

#### 1) Dedicated Co-Processor-Based HSMs

In a dedicated co-processor–based architecture, the HSM is implemented as an independent processing unit within the SoC, featuring its own CPU core, local memory, and peripheral interfaces. The HSM executes security firmware independently of the main application processor and communicates with it through a well-defined command or mailbox interface.

This model offers strong isolation, as security-critical code and assets are physically separated from application software. Dedicated HSMs are well suited for implementing secure boot verification, key management, and cryptographic services that must remain trustworthy even if the host processor is fully compromised. However, this approach incurs higher silicon area and integration cost, and requires careful interface design to avoid introducing new attack surfaces through inter-processor communication.

#### 2) Security Islands and Isolated Execution Regions

Security Island architectures implement the HSM as a logically isolated subsystem within the SoC rather than a fully independent co-processor. Isolation is achieved through a combination of hardware mechanisms such as separate clock and reset domains, memory protection units, and bus firewalls. The HSM may share certain resources, such as cryptographic accelerators or memory blocks, with the rest of the SoC while maintaining restricted access policies.

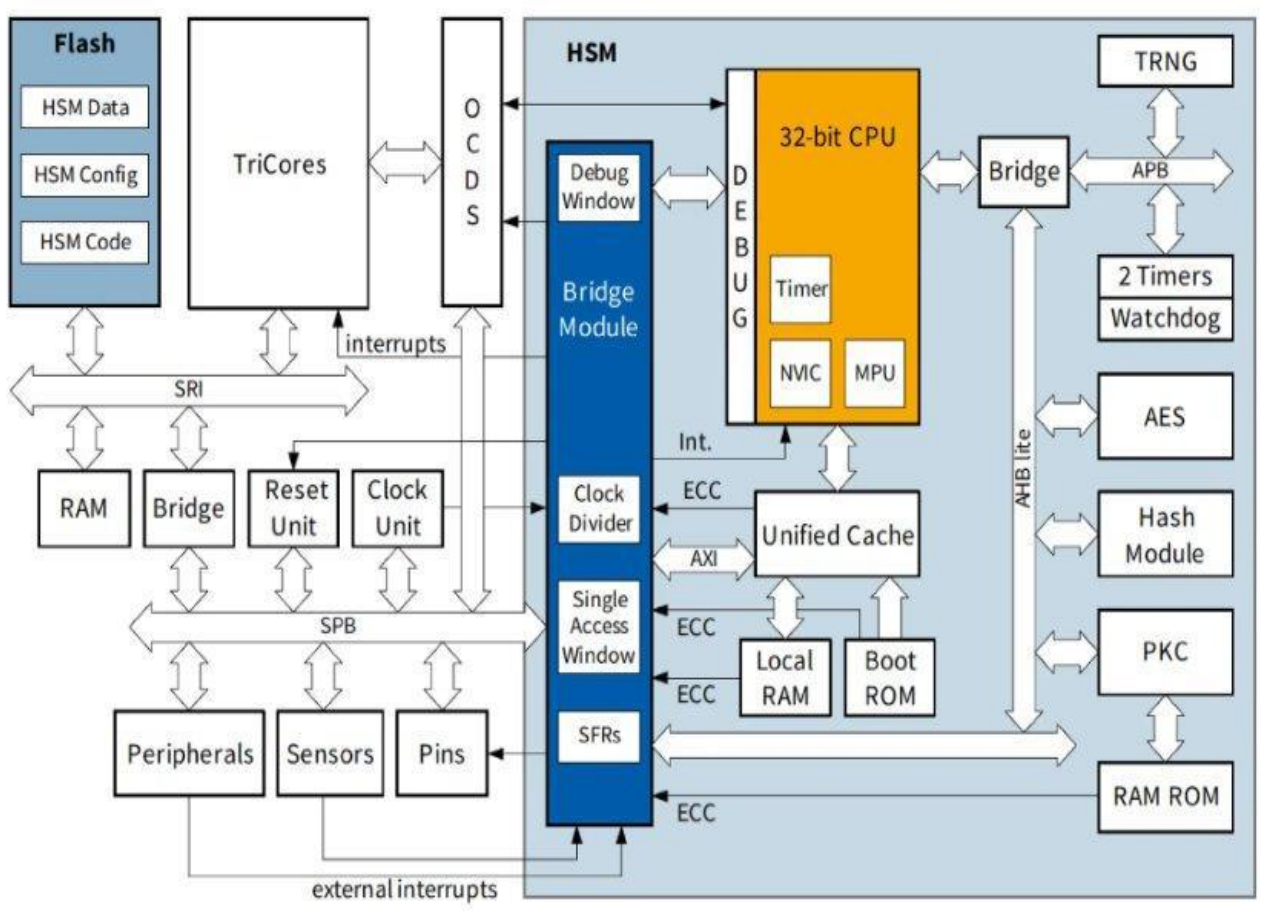

**FIGURE 1. AURIX TC3XX Architecture.**

This integration model offers a balance between isolation and cost efficiency. By reusing existing SoC components, security islands reduce silicon overhead while still providing a reasonably strong trust boundary. However, the shared nature of some resources introduces additional complexity in verification and increases the risk of cross-domain attacks if isolation mechanisms are misconfigured or compromised.

#### 3) Hardware-assisted firmware-based HSMs

In hardware-assisted firmware-based designs, the HSM functionality is primarily implemented in firmware running on a general-purpose core, with security enforced through hardware features such as privileged execution modes, memory protection, and cryptographic accelerators. Rather than being a separate processing entity, the HSM operates as a constrained execution environment within the main processor complex.

This approach maximizes flexibility and reduces hardware cost, making it attractive for lower-end ECUs or platforms with tight area constraints. However, the security guarantees of firmware-based HSMs are inherently weaker than those of more isolated designs, as they rely on correct configuration and enforcement of shared hardware resources. Consequently, such architectures are more sensitive to software vulnerabilities and configuration errors, particularly in early boot stages.

### B. Isolation Mechanisms and Trust Boundaries

Isolation is the defining characteristic of an automotive HSM and the primary determinant of its security strength. HSM architectures employ a combination of physical and logical isolation mechanisms to protect security-critical assets from unauthorized access. These mechanisms include dedicated execution cores, isolated memory regions, bus-level access controls, and privileged instruction sets.

Trust boundaries delineate which components are considered part of the trusted computing base (TCB) and which are treated as potentially untrusted. A smaller and more tightly scoped TCB generally results in a reduced attack surface and easier security verification. However, narrowing the trust boundary often requires additional hardware resources or restricts functionality. Architectural decisions regarding isolation therefore directly influence both security assurance and system complexity.

### C. Cryptographic Acceleration and Service Interfaces

Automotive HSMs typically incorporate hardware accelerators for commonly used cryptographic primitives to meet performance and real-time requirements. These accelerators offload computationally intensive operations such as symmetric encryption, digital signatures, and hash computations from the main CPU, enabling deterministic execution and reduced latency.

Access to cryptographic services is usually mediated through a defined service interface that abstracts HSM internals from application software. This interface may take the form of command queues, memory-mapped registers, or message-based protocols. While richer interfaces improve usability and flexibility, they also expand the exposed attack surface. Consequently, interface simplicity and strict input validation are critical considerations in HSM design.

### D. Flexibility vs. Security: Programmability and Updateability

Programmability and updateability are double-edged swords in automotive HSM design. On one hand, programmable HSM firmware allows manufacturers to add features, fix vulnerabilities, and adapt to evolving cryptographic requirements over the vehicle's lifetime. On the other hand, increased programmability enlarges the attack surface and complicates formal verification and certification.

Many automotive HSMs adopt a hybrid approach, combining immutable hardware or ROM-based components that establish the root of trust with updateable firmware layers that implement higher-level services. Secure update mechanisms, version control, and rollback protection are essential to prevent unauthorized modification of HSM behavior. The degree of allowed updateability is therefore a critical architectural decision that reflects a tradeoff between long-term maintainability and immediate security assurance.

### E. Summary of Architectural Tradeoffs

The architectural design of an automotive HSM reflects a series of interconnected tradeoffs. Stronger isolation generally improves resistance against powerful adversaries but increases cost and integration complexity. Greater flexibility supports long vehicle lifecycles and evolving requirements but expands the trusted computing base and attack surface. Similarly, richer service interfaces improve usability at the expense of increased exposure.

Rather than a one-size-fits-all solution, automotive HSM architectures occupy a spectrum defined by isolation strength, performance determinism, cost, and adaptability. Understanding these tradeoffs is essential for system architects seeking to align security objectives with the practical constraints of automotive platforms. In the following sections, these architectural considerations are examined in the context of secure boot, key management, and threat models relevant to modern vehicles.

## IV. CURRENT USE CASES OF HSMS IN AUTOMOTIVE ECUS

Automotive HSMs are no longer experimental components instead they are a standard part of security architectures in many production ECUs. In today's vehicles, HSMs are primarily used to establish trust during system startup, protect long-term cryptographic assets, and support authenticated communication and diagnostic access. Rather than acting as general-purpose security engines, current automotive HSM deployments focus on a limited set of well-defined functions that align with real-time, cost, and lifecycle constraints. This section summarizes the most common HSM use cases observed in production automotive systems and outlines the practical boundaries within which these modules are typically employed.

### A. Secure Boot Support and Boot-Time Verification

Secure boot is one of the earliest and most widely adopted use cases for automotive HSMs. During ECU startup, the HSM is used to verify the authenticity and integrity of firmware components before execution proceeds beyond the initial boot stages. Digital signatures on bootloaders, operating systems, or application images are validated using cryptographic keys that are protected by the HSM and cannot be extracted by application software.

In many automotive SoCs, immutable boot ROM code initiates the boot process and invokes cryptographic services provided by HSM. This establishes a chain of trust rooted in hardware, ensuring that only authorized software is allowed to execute, even if external flash memory has been modified. Unlike confidentiality-oriented designs, automotive secure boot typically focuses on authenticity and integrity, allowing firmware to remain unencrypted while still preventing unauthorized code execution.

This model has proven effective in mitigating firmware tampering and persistent compromise, and it has become a foundational requirement in modern automotive security architectures, particularly for ECUs involved in safety-critical functions or software update workflows [1], [4].

### B. Secure Key Storage and Lifecycle Management

Another core function of automotive HSMs is the secure storage and management of cryptographic keys. Keys used for secure boot verification, in-vehicle communication, diagnostic authentication, and software updates are typically provisioned into the HSM during manufacturing or early lifecycle stages and are designated as non-exportable. This ensures that sensitive cryptographic material remains protected even if the main processor or operating system is compromised.

Beyond static storage, many HSMs support basic key lifecycle operations, including key derivation, usage restrictions, and controlled revocation. In practice, however, key lifecycle management in production systems is constrained by limited secure memory and fixed key slot structures. As a result, OEMs and suppliers must carefully design provisioning workflows and key hierarchies to fit within these constraints while still meeting security and regulatory requirements [2], [3].

### C. Cryptographic Services for In-Vehicle Communication

Automotive HSMs commonly provide cryptographic services to support authenticated in-vehicle communication, including message authentication, hashing, symmetric encryption, and random number generation. Hardware acceleration enables these operations to meet strict timing and performance requirements. In production deployments, HSMs are often used to generate or verify message authentication codes using session keys derived from long-term secrets stored in hardware. Offloading cryptographic operations to the HSM reduces key exposure in system memory and improves execution determinism compared to software-only implementations. Access to these services is typically provided through constrained interfaces that prioritize predictability over protocol flexibility.

### D. Support for Secure Diagnostics and Vehicle Identity

Secure diagnostic access is another important application of automotive HSMs. Modern diagnostic protocols increasingly rely on cryptographic authentication to restrict access to sensitive ECU functions such as reprogramming, calibration, or parameter modification. The HSM is commonly used to store diagnostic credentials and to perform authentication operations without exposing secrets to application software. In addition to diagnostics, HSMs often store identity-related material such as certificates or unique identifiers that are used to establish ECU or vehicle identity within backend systems. Hardware-backed identity protection reduces the risk of cloning or impersonation attacks that could otherwise be enabled through firmware extraction or memory analysis. This capability is particularly relevant for software update authorization, backend communication, and supply-chain traceability [2]. In many current deployments, identity credentials are treated as relatively static assets, with limited support for renewal or migration over the vehicle's lifetime. This reflects both architectural limitations and the complexity of coordinating identity management across OEM, supplier, and service infrastructures.

### E. Limitations of HSM Usage in Current Production Systems

Despite widespread deployment, HSM usage in production systems is often limited to a narrow set of functions, primarily secure boot and basic key storage. Resource constraints such as limited secure memory, bounded processing capacity, and real-time latency requirements frequently restrict broader use of HSM services. As a result, some cryptographic operations remain outside the HSM, increasing the overall trusted computing base. Fixed assumptions about key sizes, algorithms, and verification latency further constrain long-term adaptability. In addition, the security benefits of an HSM depend heavily on correct system integration; misconfigured access controls or service interfaces can significantly weaken intended protections. These limitations motivate a deeper examination of attacker models and threat assumptions, which is addressed in the next section.

## V. THREAT MODELS FOR AUTOMOTIVE ECUs

A defensible automotive security architecture must begin with explicit and conservative assumptions. The attacker classes considered in this work, summarized in Appendix A, represent realistic threat actors that may interact with automotive ECUs throughout the vehicle lifecycle. These range from remote adversaries exploiting network connectivity to physically capable attackers performing offline analysis of extracted hardware. Software vulnerabilities must be assumed inevitable over a vehicle's lifetime that may exceed 15 years. Complex protocol stacks, third-party components, and evolving connectivity requirements make remote code execution a realistic scenario for at least one ECU within a vehicle network. In addition, physical access to vehicles cannot be prevented. Attackers may obtain temporary or permanent access, extract ECUs, and perform offline analysis without time pressure. Backend and supply chain environments also represent high value targets. Although strong process controls reduce the likelihood of compromise, it cannot be categorically excluded. Consequently, critical trust anchors must be hardware enforced and minimally dependent on mutable software state.

Under these assumptions, realistic attacker capabilities vary by class but should not be underestimated. Remote attackers may exploit memory corruption vulnerabilities, logic flaws, or configuration errors to achieve arbitrary code execution within application contexts. From there they may attempt privilege escalation, lateral movement to adjacent ECUs, or manipulation of update workflows. Local logical attackers may leverage diagnostic services, misconfigured debug interfaces, or service modes to reflash firmware or modify security relevant configuration. Physical attackers, even without invasive semiconductor modification, can perform noninvasive probing, voltage or clock glitching, and basic side channel measurements using commercially available equipment. Highly sophisticated invasive attacks remain possible but are typically constrained by cost and scalability considerations. These attacker capabilities ultimately target a defined set of security relevant assets within the ECU and the broader vehicle ecosystem. The primary assets considered in this work, including boot firmware, cryptographic keys, firmware integrity, configuration data, and device identity, are summarized in Appendix B.

Accordingly, security mechanisms such as secure boot, key storage, and lifecycle controls must remain robust even if the main processor is compromised and must maintain security guarantees under physical access conditions. Long term cryptographic keys should therefore be treated as high value assets and protected against extraction, including during offline attack scenarios. At the same time, defenses must remain economically viable for mass produced ECUs. This requires a calibrated approach that prioritizes resistance to scalable real-world attacks while acknowledging the diminishing returns of defending against highly specialized laboratory adversaries.

## VI. SECURE BOOT, HARDWARE ROOTS OF TRUST, AND HSM INTEGRATION IN AUTOMOTIVE SOCS

Secure boot provides the foundation of software integrity in modern automotive ECUs. As vehicles transition toward software-defined architectures with persistent connectivity and frequent updates, ensuring that only authenticated software executes has become a baseline requirement. Secure boot establishes this assurance through a hardware-anchored chain of trust that validates each stage of initialization before execution.

### A. Secure Boot Fundamentals and Chain-of-Trust Models

Execution begins from an immutable region, typically an on-chip boot ROM, which verifies the integrity and authenticity of the next-stage bootloader using cryptographic signatures. Each subsequent stage verifies the next, forming a hierarchical chain of trust from hardware to application software. Two principal models are employed: verification-based chains of trust and measurement-based approaches [6], [7]. Verification models authenticate each stage prior to execution and are widely adopted in automotive systems due to deterministic behavior aligned with safety requirements. Measurement-based models record integrity values for later validation and are more common in general-purpose computing environments. The robustness of the chain ultimately depends on protecting verification keys and preserving immutability of the first executed instruction. Compromise at this stage enables persistent system control.

### B. Hardware Roots of Trust in Automotive SoCs

A hardware root of trust consists of immutable or hardware-protected elements anchoring system security, typically including:

A. Mask-programmed boot ROM,

B. Hardware-protected key storage (e.g., OTP or secure NVM),

C. Privileged execution modes and memory isolation

Immutability and tamper resistance are essential. If early verification logic or embedded keys can be modified or extracted, the secure boot mechanism collapses. As noted in foundational work, practical compromise often stems from weak boundary enforcement rather than broken cryptography [8]. Automotive constraints (e.g. cost, long lifecycles, and real-time determinism) have driven tight integration of these trust anchors within SoCs rather than reliance on discrete security components.

### C. HSM-Assisted Secure Boot Architectures

In production systems, secure boot robustness depends heavily on how Hardware Security Modules (HSMs) are integrated within the SoC. Instead of embedding full verification logic in boot ROM, many platforms delegate signature verification and key protection to an integrated HSM.

Two common patterns emerge:

A. HSM as verification engine, where boot ROM invokes HSM services using protected keys.

B. HSM as trust anchor, where the HSM contains root keys and enforces firmware acceptance policies.

Delegation reduces ROM complexity while strengthening key isolation. However, when the HSM becomes the trust anchor, its own integrity and configuration become security-critical.

### D. Key Protection and Cryptographic Offloading

boot depends on hardware-protected storage of verification keys. Storing keys in external flash exposes them to extraction and modification. Modern HSMs mitigate this through protected key slots accessible only via controlled interfaces, as formalized in AUTOSAR Secure Hardware Extensions (SHE) [9]. HSMs typically integrate cryptographic accelerators to ensure bounded verification latency. Deterministic execution is critical in automotive environments; therefore, strict interface controls must ensure that verification requests originate from authorized boot stages and that results cannot be bypassed or spoofed. The CPU–HSM interface boundary is thus as important as the cryptographic primitive itself.

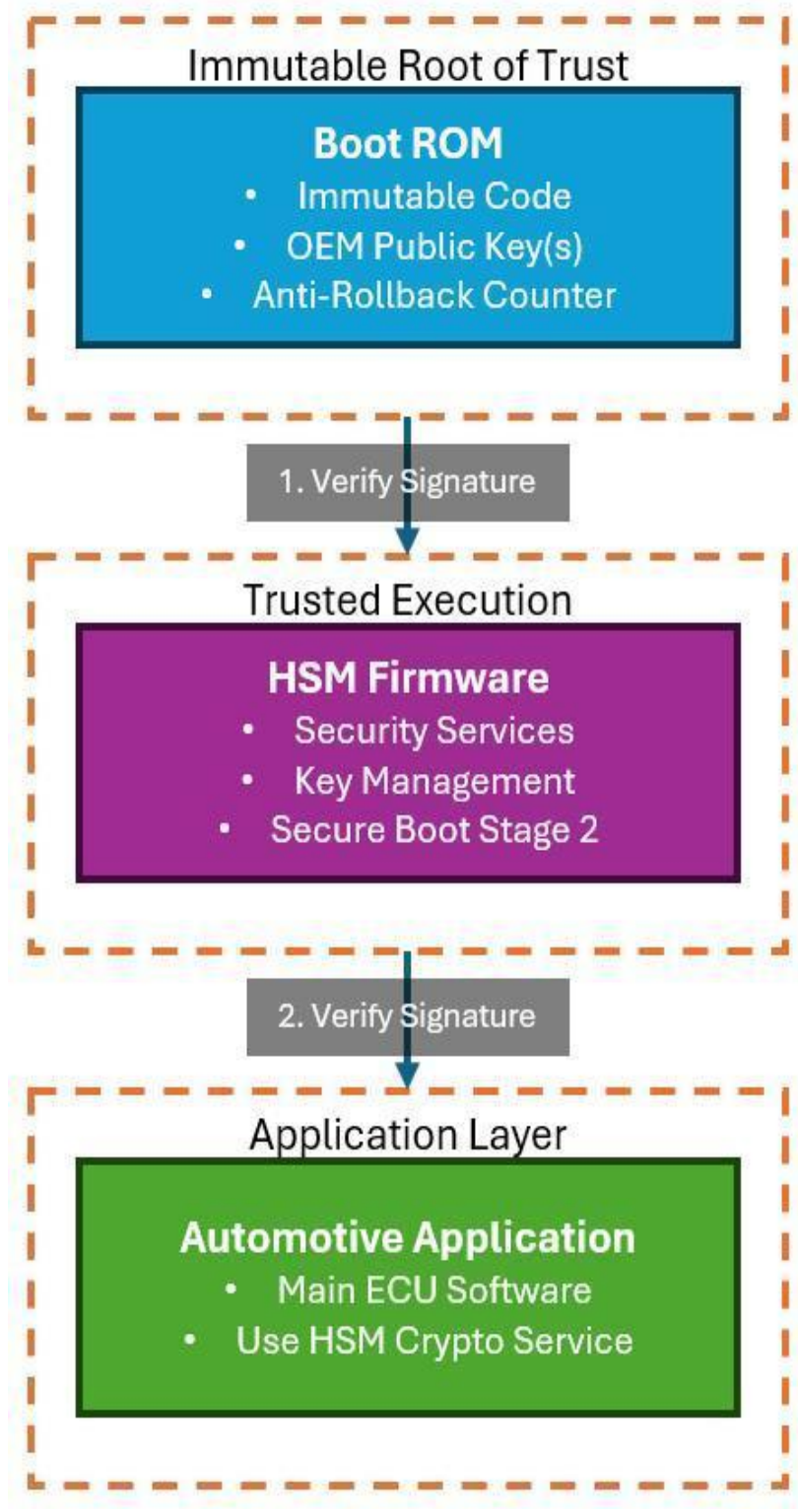

**FIGURE 2. HSM-Assisted Secure Boot Flow.**

### *E. Architectural Variants and Security Implications*

Isolation strength varies by implementation:

A. Dedicated coprocessor HSMs — provide strong separation but increase cost.

B. Security island architectures — balance isolation and integration efficiency.

C. Firmware-based HSM designs — improve flexibility but expand the trusted computing base.

Secure boot effectiveness depends not only on cryptographic correctness but on careful trust placement and isolation enforcement across these variants.

### *F. Failure Modes and Architectural Limits*

Without hardware enforcement, secure boot remains vulnerable to bootloader replacement, key extraction, execution redirection, and rollback attacks. Automotive security research has demonstrated that early-stage compromise enables persistent ECU control. Even HSM-assisted designs embed assumptions regarding key sizes, verification latency, and protected memory allocation. These parameters influence accelerator sizing, key-slot provisioning, and ROM budgeting. If cryptographic requirements evolve, fixed hardware limits may constrain adaptability. Secure boot architectures must therefore incorporate sufficient abstraction and capacity to remain resilient over long vehicle lifecycles.

## VII. SECURE STORAGE AND KEY MANAGEMENT

Secure storage and key management form the core of an automotive Hardware Security Module (HSM). In modern ECUs, cryptographic keys anchor secure boot, software authenticity, in-vehicle communication protection, and backend authentication. Because automotive systems operate in physically accessible and long-lived environments, key material must remain protected against invasive hardware attacks, firmware compromise, and lifecycle transitions such as manufacturing, servicing, and decommissioning. Automotive HSM architectures therefore combine hardware isolation, tightly controlled execution contexts, and lifecycle-aware provisioning mechanisms to ensure that sensitive material is never exposed to untrusted software or external interfaces.[14]

### *A. Hardware-Backed Key Storage Models*

Hardware-backed key storage ensures that cryptographic keys are generated, stored, and used within a trusted boundary resistant to both software and physical attacks. Automotive HSMs typically implement one or more of the following models:

#### 1) EMBEDDED NON-VOLATILE SECURE MEMORY

Dedicated on-chip NVM, shielded by access control logic and accessible only from within the HSM execution environment. This approach provides strong isolation and low latency, but capacity is limited.

#### 2) DERIVED KEY MODELS USING HARDWARE ROOTS

Instead of permanently storing all keys, a device-unique root secret — often injected at manufacturing — is stored in tamper-resistant memory. Operational keys are derived at runtime using key derivation functions. This reduces NVM requirements and limits exposure of long-term keys.

#### 3) PHYSICALLY UNCLONABLE FUNCTION (PUF)-BASED STORAGE

In some SoCs, secrets are reconstructed from intrinsic silicon characteristics rather than stored directly. PUF-based designs reduce static key presence in memory, though they introduce reliability and environmental compensation considerations.

#### 4) EXTERNAL SECURE ELEMENTS

Certain architectures use discrete secure elements connected via secure interfaces. While this improves physical isolation, it increases system cost and integration complexity compared to tightly integrated HSMs.

### *B. Key Provisioning, Usage, and Revocation*

Automotive ECUs operate within complex supply chains involving semiconductor vendors, Tier-1 suppliers, and OEMs. As a result, key management must support secure provisioning across multiple trust domains.

#### 1) PROVISIONING

Initial root keys are typically injected during wafer test or final device personalization. OEM-specific keys may be provisioned later in the manufacturing chain. Secure provisioning requires authenticated channels, anti-replay protection, and traceability to prevent overproduction or key cloning.

#### 2) CONTROLLED KEY USAGE

Keys stored in the HSM are bound to usage policies enforced in hardware. For example, a boot verification key may only be used for signature verification, while a communication key may be restricted to specific cryptographic algorithms. Raw key export is generally prohibited.

#### 3) REVOCATION AND UPDATE

Given vehicle lifetimes exceeding 10–15 years, cryptographic material must support revocation. This includes:

1) Certificate revocation lists or embedded trust anchors,
2) Monotonic counters to invalidate outdated firmware,
3) Secure key rotation mechanisms compatible with over-the-air (OTA) updates

Hardware-enforced lifecycle states (e.g., development, production, service, decommissioned) restrict which keys are usable at any time, reducing exposure during debugging or maintenance.

### C. Protection Against Firmware Extraction and Rollback

Firmware protection relies on the integrity guarantees provided by the secure boot chain. At reset, immutable boot ROM verifies the authenticity of the next-stage firmware using keys anchored in the HSM. Only verified code gains execution privileges.

To prevent firmware extraction:

1) Debug interfaces are permanently or conditionally locked.
2) Memory regions are encrypted or access-controlled.
3) Cryptographic accelerators refuse operations outside authorized contexts.

Rollback protection is critical in safety-relevant automotive systems. Attackers may attempt to install an older, vulnerable firmware version with valid signatures. HSMs mitigate this via:

1) Secure version counters stored in tamper-resistant memory,
2) Anti-rollback fuses,
3) Firmware version binding in signature metadata.

These measures ensure that only firmware meeting or exceeding a defined security baseline can execute.

### D. OEM and Supplier Trust Boundaries

Automotive security architectures must reflect layered trust relationships. Semiconductor vendors implement the HSM hardware and root mechanisms. Tier-1 suppliers integrate ECUs and application software. OEMs define vehicle-level security policies and backend trust anchors.

Key separation mechanisms enforce these boundaries:

1) Distinct root keys for silicon vendor and OEM.
2) Hardware-enforced partitioning of key slots.
3) Access control lists tied to authenticated software identities.
4) Secure boot chains that delegate trust without exposing root secrets.

For example, the silicon vendor may control immutable boot ROM verification, while the OEM controls firmware signing keys for field updates. Proper boundary definition prevents a compromised supplier from accessing OEM private material and vice versa.

Clear delineation of trust domains also simplifies compliance with automotive cybersecurity standards such as ISO/SAE 21434 and supports alignment with broader security frameworks like ISO/IEC 27001 where applicable to backend integration.

### E. Scalability of Hardware Key Storage Under Evolving Cryptographic Primitives

Automotive ECUs must remain secure across decades, during which cryptographic standards evolve. Migration from RSA to elliptic-curve cryptography (ECC), and potentially to post-quantum algorithms, introduces significant storage and performance implications.

Key challenges include:

A. Increased key sizes: Post-quantum schemes may require substantially larger public keys and signatures, stressing limited secure NVM capacity.
B. Algorithm agility: HSM architectures must support multiple algorithms concurrently to enable phased transitions. This requires flexible key metadata structures and programmable policy engines.
C. Performance constraints: Real-time automotive workloads limit acceptable cryptographic latency. Larger primitives increase computational overhead and memory bandwidth demands.
D. Secure migration paths: Hybrid signature models (e.g., classical + post-quantum) may be necessary during transition phases. Hardware support must enable coexistence without exposing downgrade vulnerabilities.

Scalable key storage architectures therefore favor abstraction layers that decouple key identity from underlying primitive representation. Future automotive HSMs will likely integrate configurable cryptographic accelerators and modular firmware architectures to accommodate evolving standards without silicon redesign.

In summary, secure storage and key management within automotive HSMs must balance strong hardware isolation, lifecycle-aware provisioning, rollback resistance, supply-chain trust separation, and forward-looking cryptographic agility. These elements collectively establish a resilient hardware root of trust capable of sustaining long-term vehicle security.

## VIII. SECURE EXECUTION AND SOFTWARE ISOLATION

Beyond secure storage, automotive HSMs provide an isolated execution environment for security-critical services. In modern ECUs, cryptographic verification, key derivation, secure state transitions, and parts of the secure boot chain execute within the HSM boundary rather than in the main application processor. This architectural separation reduces the attack surface of safety- and mission-critical functions while maintaining compatibility with real-time automotive workloads. Secure execution in automotive SoCs is typically achieved through a tightly coupled coprocessor model: the HSM contains its own CPU core, internal memory (RAM/ROM), cryptographic accelerators, and access control logic. The host processor interacts with the HSM through a restricted command interface, ensuring that sensitive operations are mediated and policy-enforced in hardware.

### A. Isolated Execution of Security-Critical Services

The HSM executes a dedicated firmware image responsible for:

1) Secure boot verification and chain-of-trust enforcement
2) Cryptographic primitives (e.g., signature verification, MAC generation, encryption/decryption)
3) Secure key derivation and wrapping
4) Lifecycle state management
5) Secure debug authorization

Isolation is enforced at multiple levels:

1) Memory isolation: HSM-internal RAM and ROM are not directly addressable by the host CPU.
2) Peripheral isolation: Cryptographic accelerators are accessible only via HSM firmware.
3) Privilege separation: Host software cannot execute arbitrary code inside the HSM.

This separation ensures that even if the main application processor is compromised (via remote exploitation or memory corruption) the attacker cannot directly extract keys or bypass secure boot policies. In production ECUs, the HSM firmware itself is authenticated during system startup and may be stored in immutable ROM or securely updatable flash with strict signature verification requirements.

### B. Trusted Computing Base (TCB) Minimization

A core design objective of automotive HSMs is minimizing the Trusted Computing Base (TCB), the set of components that must function correctly to guarantee security.

TCB minimization improves resilience by:

1) Reducing the amount of code that must be verified and maintained securely
2) Limiting exposure to software vulnerabilities
3) Simplifying formal verification and certification efforts

In an HSM-centric design, the TCB typically includes:

1) Boot ROM code
2) HSM firmware
3) Hardware-enforced access control mechanisms
4) Cryptographic accelerators

Application-layer software, operating systems (e.g., AUTOSAR stacks), and communication middleware remain outside the security-critical TCB. By restricting key handling and trust decisions to a compact HSM firmware, designers reduce systemic risk. However, TCB minimization must be balanced against flexibility. Excessive reliance on host software for policy enforcement re-expands the effective TCB, while overly complex HSM firmware increases verification difficulty.

### C. Interaction Between Application Software and HSM Services

The host processor communicates with the HSM through a command-response interface, often implemented via memory-mapped registers, mailboxes, or shared buffers. All requests (such as “verify signature,” “derive key,” or “generate random number”) are validated against hardware-enforced policies before execution.

Typical interaction properties include:

1) Command whitelisting: Only predefined service identifiers are accepted.
2) Parameter validation: Buffer sizes, memory regions, and key identifiers are checked before use.
3) Access control binding: Keys are bound to usage permissions stored in secure metadata.
4) Asynchronous execution: For computationally heavy operations, the HSM may signal completion via interrupts.

This model prevents raw key material from leaving the HSM boundary. Instead, cryptographic operations are performed internally, and only non-sensitive outputs (e.g., signatures or verification results) are returned. Careful API design is critical: overly flexible command sets increase attack surface, while rigid interfaces may hinder future cryptographic agility.

### D. Security Guarantees and Limitations of HSM-Based Execution

HSM-based isolation provides strong guarantees:

1) Protection of key material against software compromise
2) Enforcement of secure boot and anti-rollback policies
3) Hardware-mediated lifecycle state transitions
4) Resistance to many classes of remote exploitation

However, its guarantees are not absolute.

Limitations include:

1) Side-channel leakage: Power analysis or electromagnetic probing may target cryptographic accelerators.
2) Fault injection attacks: Glitching or voltage manipulation may attempt to bypass verification logic.
3) Shared resource leakage: Cache or bus contention between host and HSM can create indirect channels.
4) Firmware complexity risk: Bugs within HSM firmware affect the entire trust chain.

Mitigations often include hardware countermeasures, redundancy checks, constant-time cryptographic implementations, and secure state machines. Security certification efforts, such as compliance with ISO/SAE 21434 and alignment with functional safety frameworks like ISO 26262, frequently influence architectural decisions. While these standards do not prescribe specific HSM designs, they shape expectations around isolation strength, traceability, and risk assessment. Ultimately, the HSM protects against a well-defined adversary model; it cannot compensate for systemic design flaws in higher-level vehicle architectures.

### E. Long-Latency Cryptographic Operations and Real-Time Implications

Automotive systems operate under strict real-time constraints. Boot-time deadlines, communication latency requirements (e.g., for secure in-vehicle messaging), and safety-critical control loops limit acceptable cryptographic overhead. Long-latency operations may introduce measurable delays, such as:

1) Asymmetric signature verification (e.g., large RSA keys)
2) Post-quantum signature validation
3) Bulk secure firmware decryption
4) Secure key provisioning sequences

Key architectural tradeoffs include:

1) Hardware acceleration vs. silicon cost: Dedicated accelerators reduce latency but increase die area.
2) Blocking vs. non-blocking APIs: Asynchronous HSM interfaces allow the host CPU to continue execution while cryptographic tasks complete.
3) Boot-time optimization: Parallel verification stages can shorten secure boot chains.
4) Deterministic timing: Safety-critical systems require bounded worst-case execution times, not merely average performance.

As cryptographic primitives evolve, particularly with the anticipated integration of post-quantum schemes, signature sizes and computation times may increase significantly. Without careful architectural planning, this can conflict with real-time startup and update constraints. Future automotive HSM designs will likely incorporate configurable acceleration pipelines, multi-core security coprocessors, or staged verification mechanisms to maintain deterministic performance while preserving strong isolation guarantees. In summary, secure execution within automotive HSMs strengthens isolation, reduces effective TCB size, and enforces policy at the hardware boundary. However, achieving robust protection while maintaining real-time determinism requires careful co-design of hardware architecture, firmware complexity, and system-level scheduling.

## IX. SOFTWARE SIGNING AND SECURE UPDATE MECHANISMS

Secure boot establishes initial trust at power-on, but maintaining that trust over the vehicle lifecycle requires robust software signing and secure update mechanisms. Modern vehicles routinely receive firmware updates to patch vulnerabilities, introduce features, and meet regulatory requirements. As a result, the integrity of the software signing process and the correctness of update validation mechanisms are as critical as the boot chain itself.

### A. Software Signing Workflows in Automotive Systems

Automotive software signing typically follows a hierarchical trust model. A root key, controlled by the OEM, authorizes one or more intermediate keys, which in turn sign firmware images. This layered approach enables separation of responsibilities between OEMs, suppliers, and manufacturing environments while preserving centralized trust control.
Firmware packages generally include:

1) The executable image,
2) A manifest containing metadata (version, target ECU, rollback counter),
3) A digital signature covering both payload and metadata.

This signing workflow aligns with industry recommendations for secure update frameworks, where authenticity, integrity, and version control are enforced prior to installation [13]. In production environments, signing keys are commonly protected within enterprise-grade HSMs to prevent unauthorized firmware generation.

The security of the entire vehicle fleet depends on strict control of private signing keys. Compromise at this level can enable large-scale deployment of malicious but validly signed firmware. Public Key Infrastructure (PKI) underpins this trust hierarchy by enabling scalable key distribution, certificate validation, and revocation across vehicle fleets, ensuring that firmware authenticity can be verified even in the presence of distributed development and supply-chain complexity [10].

### B. HSM Role in Firmware Authentication and Validation

On the vehicle side, HSMs provide hardware-protected storage for public verification keys and enforce controlled access to signature verification services. During both boot-time and update-time validation, the main processor invokes HSM services to authenticate firmware images before execution.
This architecture ensures:

1) Verification keys are inaccessible to application software,
2) Signature verification cannot be bypassed through memory modification,
3) Cryptographic operations are executed within an isolated environment.

AUTOSAR Secure Hardware Extensions (SHE) exemplifies this model by defining fixed key slots and restricting direct key access [9]. Hardware-backed verification significantly reduces the attack surface compared to software-only validation approaches.

### C. Secure Boot and OTA Update Interactions

Over-the-air (OTA) updates introduce additional complexity beyond initial boot verification. The typical OTA flow involves:

1) Secure download of update package,
2) Storage in a staging partition,
3) Signature verification and policy checks,
4) Activation on next reboot,
5) Re-verification during secure boot

If any step is improperly implemented, attackers may exploit the update mechanism to install malicious or downgraded firmware. Anti-rollback protections, such as monotonic counters stored in protected memory, are essential. Prior work on automotive update security has shown that weaknesses in update validation or rollback enforcement can enable persistent ECU compromise [13]. Secure boot and OTA validation must operate as a unified system. Boot-time verification alone cannot prevent compromise if the update pipeline allows malicious code to be accepted as valid.

### D. Common Implementation Pitfalls and Attack Vectors

Despite widespread adoption of digital signing, practical implementation weaknesses continue to emerge. Common issues include:

1) Reuse of signing keys across multiple product lines,
2) Insecure provisioning of verification keys during manufacturing,
3) Debug interfaces left enabled or unprotected in production ECUs,
4) Failure to verify all update metadata,
5) Time-of-check to time-of-use (TOCTOU) vulnerabilities,
6) Insufficient rollback protection.

High-profile automotive security research has demonstrated that exploitation of update or firmware validation weaknesses can lead to persistent control over safety-critical ECUs [12], [13]. These cases illustrate that cryptographic signatures alone are insufficient; architectural enforcement and lifecycle governance are equally important.

### E. Signature Size, Bandwidth, and Verification Overhead in OTA Pipelines

Software signing introduces computational, storage, and communication overhead that must be carefully managed in resource-constrained ECUs. Larger signatures and manifests increase OTA payload size, extend download windows, and require additional flash staging capacity prior to activation. On-device verification latency correspondingly grows, potentially affecting boot timing and update availability windows. In HSM-assisted architectures, these effects are amplified by fixed accelerator capabilities, bounded protected memory, and constrained key-slot provisioning. Verification services must remain deterministic and isolated, and increased signature size or computational complexity can stress both execution timing and secure storage allocation. Consequently, scalable update design is not merely a cryptographic concern, but an architectural constraint that directly influences HSM integration.

## X. DISCUSSION: DESIGN TRADEOFFS AND OPEN CHALLENGES

The preceding sections examined how secure boot, hardware roots of trust, HSM integration, and secure update mechanisms collectively establish software integrity in automotive ECUs. However, these mechanisms do not operate in isolation. They are shaped by practical constraints, lifecycle requirements, and ecosystem fragmentation. This section discusses the broader architectural tradeoffs and open challenges that influence long-term resilience of automotive hardware security.

### A. Security vs. Performance and Real-Time Constraints

Automotive systems operate under strict real-time constraints that limit the computational and memory overhead tolerated during boot and update operations. While strong cryptographic verification enhances security assurance, it introduces latency, increased memory footprint, and additional inter-processor communication between the main CPU and the HSM. Secure boot verification must be completed within bounded startup windows, particularly in safety-critical ECUs where availability requirements are tightly defined. Offloading verification to hardware accelerators mitigates some latency, yet accelerator throughput and protected memory capacity remain finite. In resource-constrained controllers, prolonged verification or large manifests may impact deterministic startup behavior. Embedded cryptographic hardware studies have consistently highlighted the tension between isolation strength and real-time determinism. Architectural decisions must therefore balance security margins against predictable execution timing. Over-dimensioning security services increases cost and silicon area whereas under-dimensioning risks future infeasibility.

### B. Updateability vs. Attack Surface Expansion

Long vehicle lifecycles demand updateable security mechanisms. Firmware-based HSM services allow vulnerability remediation and feature evolution, but they also enlarge the trusted computing base. Each additional update path, debug interface, or service endpoint expands the attack surface. Rigid hardware implementations offer stronger assurance but reduce adaptability over a vehicle's operational lifetime. Conversely, highly programmable security components improve flexibility while introducing complexity in validation, testing, and certification. This tradeoff is particularly acute in software-defined vehicle platforms where continuous integration pipelines intersect with embedded trust anchors. Experience across embedded domains suggests that compromise frequently arises from boundary misconfiguration rather than cryptographic weakness. Thus, updateability must be designed with strict governance, limited privilege exposure, and clear separation between operational and security domains.

### C. Standardization Gaps and Interoperability Challenges

Although standards such as AUTOSAR Secure Hardware Extensions provide baseline guidance for key storage and cryptographic services, substantial variability remains across SoC vendors and ECU implementations. HSM capabilities differ in terms of key slot capacity, supported primitives, protected memory allocation, and service interfaces. In multi-supplier vehicle architectures, inconsistent trust models and provisioning workflows can complicate integration. Differences between AUTOSAR Classic and Adaptive environments further fragment implementation approaches. As vehicle platforms evolve toward centralized and zonal architectures, ensuring interoperability of hardware-rooted trust mechanisms across heterogeneous domains becomes increasingly challenging. Standardization efforts currently focus primarily on interface definitions rather than architectural requirements for scalable roots of trust. This leaves OEMs responsible for harmonizing disparate security assumptions across components.

### D. Implications for Future Automotive SoC Designs

Emerging zonal and centralized compute platforms consolidate functionality previously distributed across numerous ECUs. This consolidation increases computational capability but also aggregates risk. A single high-performance domain controller may become responsible for multiple safety-critical subsystems, amplifying the importance of robust isolation and scalable HSM services. Future SoCs must anticipate increased firmware sizes, more frequent update cycles, and expanded certificate hierarchies. HSMs will need to be scaled not only in cryptographic throughput but also in protected storage and policy enforcement flexibility. Architectural partitioning strategies, memory isolation mechanisms, and secure interconnect designs will become central to maintaining containment across consolidated domains. These trends suggest that hardware-rooted trust must evolve in parallel with compute consolidation, rather than remaining a fixed-function afterthought.

### E. Cryptographic Agility as an Architectural Requirement

Current secure boot and update architectures implicitly assume stable cryptographic primitives, bounded key sizes, and predictable verification cost. These assumptions influence boot ROM allocation, HSM accelerator design, key slot provisioning, and memory budgeting. However, vehicle lifecycles routinely exceed a decade, during which cryptographic requirements may evolve. Architectural rigidity at the hardware level can constrain the ability to adopt new primitives or accommodate increased signature and key sizes. Fixed accelerators, limited protected storage, and inflexible boot ROM logic may become bottlenecks if verification requirements change. Even absent specific algorithmic shifts, forward compatibility demands that hardware trust anchors be designed with sufficient abstraction and capacity to support evolution. Cryptographic agility, therefore, should not be viewed solely as an algorithmic property but as an architectural characteristic. Designing HSM-integrated secure boot mechanisms with explicit headroom for computational growth and key management expansion is essential for long-term resilience.

## XI. FUTURE WORKS

The architectural mechanisms discussed throughout this paper established a foundation for hardware-rooted software integrity in automotive ECUs. However, evolving vehicle architectures, emerging threat models, and long lifecycle requirements introduce new challenges that warrant continued research and refinement. This section outlines key directions for future work in scalable automotive hardware security.

### A. Zonal Architectures and Centralized Compute Platforms

Automotive platforms are transitioning from distributed ECU topologies toward zonal and centralized compute architectures. In these environments, a single high-performance controller may consolidate functionality previously spread across multiple ECUs, including safety-critical and non-critical domains. This consolidation increases computational capability but also aggregates trust dependencies. Hardware Security Modules in centralized controllers must support multi-domain isolation, scalable key management, and policy enforcement across heterogeneous workloads. Research is needed to define architectural patterns that preserve strong isolation guarantees while supporting shared hardware resources. Future designs must also address secure partitioning, inter-domain communication control, and hierarchical trust delegation within consolidated SoCs.

### B. Stronger Physical Attacker Models

While much of today's automotive security architecture focuses on remote and logical adversaries, high-value vehicle platforms may increasingly face sophisticated physical attacks. Fault injections, side-channel analysis, probing, and chip-level extraction techniques challenge assumptions about hardware isolation. Future HSM implementations must balance cost constraints with enhanced tamper resistance, fault detection, and runtime integrity monitoring. Research into practical countermeasures (such as fault-resistant verification logic and lightweight tamper detection mechanisms) will be critical for maintaining assurance under stronger adversary models. As vehicle compute density increases, the economic feasibility of targeted hardware attacks may also change, requiring reevaluation of current threat assumptions.

### C. Cryptographic Agility and Post-Quantum Readiness

Automotive lifecycles frequently exceed a decade, during which cryptographic requirements may evolve. Current secure boot and HSM-assisted update architectures implicitly assume stable primitive sizes, bounded verification latency, and fixed accelerator capabilities. Future work must explore architectural mechanisms that enable cryptographic agility without compromising isolation guarantees. This includes abstraction layers between boot ROM and cryptographic accelerators, scalable protected storage for larger keys and certificates, and updateable policy enforcement logic within constrained trust boundaries. Post-quantum readiness represents one example of potential cryptographic evolution, but the broader architectural requirement is adaptability. Hardware trust anchors designed with sufficient abstraction and capacity will be better positioned to accommodate future transitions.

### D. Evolution of Automotive Hardware Roots of Trust

Hardware roots of trust are evolving alongside SoC integration trends. Emerging platforms increasingly combine secure enclaves, virtualization support, and hardware-backed isolation within high-performance compute nodes. Future research should examine how roots of trust can scale across multi-core, heterogeneous compute environments while preserving deterministic behavior and safety certification constraints. Integration of safety and security architectures will become increasingly important, particularly in consolidated zonal controllers. Architectural clarity regarding trust boundaries, hardware enforcement mechanisms, and lifecycle key governance will remain essential as automotive systems converge toward centralized software-defined platforms.

## XII. CONCLUSION

### A. Summary of Findings

Secure boot remains the foundational mechanism for establishing software integrity in automotive ECUs. However, this paper has shown that secure boot cannot be considered in isolation. Its effectiveness depends on hardware-enforced roots of trust, protected key storage, controlled cryptographic interfaces, and well-defined trust boundaries within the SoC. The integration model of the Hardware Security Module plays a decisive role in determining system robustness. Architectural decisions regarding isolation, key management, and verification delegation directly influence resilience against both logical and physical adversaries. Furthermore, secure update mechanisms extend this trust beyond initial boot, ensuring lifecycle integrity across vehicle deployments.

### B. Key Insights on HSM-Secure Boot Integration

Several architectural insights emerge:

A. Isolation boundaries often determine practical security more than cryptographic primitive selection.

B. Trust placement within the SoC significantly affects assurance guarantees.

C. Hardware-backed key protection is essential to prevent persistent compromise.

D. Lifecycle considerations, including updateability and scalability, must be accounted for at design time.

These observations emphasize that secure boot robustness is not merely a software concern but a consequence of hardware architectural decisions.

### C. Final Remarks on Scalable Automotive Hardware Security

Automotive systems are long-lived cyber-physical platforms operating in increasingly connected and software-driven environments. The durability of their security architecture depends on hardware trust anchors that are both robust and adaptable. As zonal architectures, centralized compute platforms, and evolving cryptographic requirements reshape the ecosystem, the design of HSM- integrated secure boot mechanisms will remain central to scalable automotive security. Architectural foresight, rather than reactive patching, will determine whether these systems can maintain integrity across decades of technological evolution.

# XIV. APPENDIX A

ATTACKER CLASSIFICATIONS

*Table 1 Attacker Classifications*

| Attacker Class | Access Level | Typical Entry Points / Position | Technical Capabilities | Primary Objectives | Relative Sophistication / Impact |
|---|---|---|---|---|---|
| **Remote Attacker** | No physical access; remote connectivity only | Telematics unit, cellular/Wi-Fi interfaces, OTA channels, exposed IP services, Backend systems | Software exploitation, remote code execution, network message injection, manipulation of update workflows | Install malicious firmware, gain persistence, bypass protections via software flaws | Moderate to high; highly scalable, fleet-wide impact |
| **Local Logical Attacker** | Physical vehicle access; logical interfaces only | Diagnostic port (UDS), maintenance interfaces, service bootloaders, misconfigured debug ports | Unauthorized reflashing, privilege escalation, misuse of diagnostics, parameter manipulation | Modify firmware, disable protections, escalate privileges | Moderate; direct system interaction increases effectiveness |
| **Physical Attacker** | Full physical ECU access | ECU extraction, PCB probing, direct flash access, exposed JTAG/SWD, chip-level access | Memory extraction, fault injection (voltage/clock glitching), side-channel analysis, reverse engineering, invasive attacks | Extract cryptographic keys, clone ECU, bypass secure boot | High; requires equipment and expertise, limited scalability |
| **Supply Chain Attacker** | Pre-deployment access (manufacturing, integration, logistics) | Firmware integration, key provisioning processes, third-party components, silicon fabrication | Firmware tampering before shipment, malicious key injection, hardware Trojan insertion | Persistent backdoors, large-scale compromise, IP exfiltration | Very high impact; systemic risk across product lines |
| **Insider Attacker** | Authorized internal access (OEM, Tier-1, service org.) | Signing infrastructure, backend systems, provisioning tools | Unauthorized firmware signing, key misuse, policy manipulation | Bypass cryptographic trust anchors, enable unauthorized updates | High; can circumvent technical controls if processes are weak |
| **Compromised Backend Attacker** | Control over OTA/cloud infrastructure | OTA servers, update metadata services, revocation infrastructure | Distribution of malicious updates, downgrade orchestration, fleet-wide manipulation | Fleet-wide compromise, denial of service, persistent control | Very high; centralized leverage across vehicles |
| **Adjacent ECU / In-Vehicle Attacker** | Control over one compromised ECU | CAN/Ethernet networks, gateway trust relationships | Malicious network traffic injection, lateral movement, abuse of inter-ECU trust | Escalation to high-value ECUs, cross-domain compromise | Moderate to high; depends on network segmentation strength |
| **Lifecycle / Decommissioning Attacker** | End-of-life or secondary market access | Salvaged ECUs, recycling channels, long-term offline analysis | Offline key extraction attempts, firmware reverse engineering | Device cloning, recovery of residual secrets, preparation for future attacks | Moderate to high; long time horizon, lower immediacy |

## XV. APPENDIX B

ASSETS AT RISK

*Table 2 Assets at risk*

| Asset Category | Examples | Why It Is Critical | Potential Impact if Compromised | Typical Protection Mechanisms |
|---|---|---|---|---|
| **Boot Firmware (Root of Trust Code)** | ROM bootloader, first-stage bootloader, secure boot configuration | Establishes hardware root of trust and chain of trust | Persistent compromise, secure boot bypass, downgrade attacks | Immutable ROM, signature verification, anti-rollback counters |
| **Cryptographic Keys** | Secure boot public keys, device-unique secrets, symmetric communication keys, OTA signing keys | Anchor trust relationships and secure communications | Firmware forgery, ECU cloning, backend impersonation, decryption of secure traffic | HSM-backed secure storage, key derivation, access control policies |
| **Application Firmware & Software Integrity** | Main control software, calibration binaries | Implements vehicle functionality and safety features | Malicious control behavior, unsafe actuation, persistent malware | Code signing, runtime integrity checks, memory protection |
| **Runtime Execution State** | Stack/heap memory, control-flow state, security-critical variables | Ensures correct execution flow and policy enforcement | Code injection, control-flow hijacking, privilege escalation | MPU/MMU isolation, control-flow protection, secure exception handling |
| **Configuration & Calibration Data** | Vehicle parameters, feature flags, variant coding | Defines safety limits and functional behavior | Performance manipulation, regulatory non-compliance, unsafe configurations | Authenticated storage, integrity verification, secure diagnostics |
| **Security Configuration & Policies** | Debug lock status, lifecycle state, access control lists | Governs enforcement of security mechanisms | Debug re-enablement, privilege escalation, weakened protection posture | Hardware lifecycle states, secure fuses, policy integrity checks |
| **Diagnostic Credentials** | Seed/key algorithms, workshop authentication secrets | Controls access to service and maintenance functions | Unauthorized reflashing, unlocking of restricted functions | Challenge–response authentication, rate limiting, HSM-managed secrets |
| **OTA Update Metadata** | Version counters, rollback protection values, manifest files | Ensures authenticity and freshness of updates | Downgrade attacks, installation of vulnerable firmware | Signed manifests, monotonic counters, secure storage |
| **Communication Secrets** | TLS credentials, V2X keys, gateway authentication keys | Protects in-vehicle and external communications | Message spoofing, man-in-the-middle attacks, fleet-level compromise | Hardware key isolation, secure key exchange, certificate validation |
| **Device Identity & Attestation Material** | Unique device IDs, certificates, attestation keys | Enables backend trust decisions and fleet management | ECU impersonation, counterfeit components, backend trust abuse | Hardware-protected identity keys, certificate chains, secure provisioning |
| **Intellectual Property (IP)** | Proprietary algorithms, control strategies, cryptographic implementations | Competitive differentiation and regulatory compliance | IP theft, cloning, counterfeit ECUs | Code obfuscation, memory encryption, hardware isolation |
| **Safety-Critical Control Parameters** | Torque limits, braking thresholds, ADAS configuration | Direct impact on vehicle safety | Physical safety risks, regulatory violations, liability exposure | Secure calibration workflows, authenticated updates, runtime plausibility checks |
| **Lifecycle & Provisioning Data** | Manufacturing keys, personalization data, secure state transitions | Ensures correct device state and ownership control | Backdoor insertion, improper lifecycle transitions, large-scale compromise | Secure provisioning flows, hardware state machines, audit controls |